\documentclass[superscriptaddress,prl,twocolumn,amsmath,amssymb,floatfix,showpacs,longbibliography]{revtex4-1}
\usepackage{graphicx}% Include figure files
\usepackage{bm}% bold mat
\usepackage{amssymb}
\usepackage{color}
\usepackage{epsfig}
\usepackage{subfigure}
\usepackage{hyperref}
\hypersetup{
    pdfstartview={FitH},
   pdfpagemode={None}
   }
\newcommand{\be}{\begin{equation}}
\newcommand{\ee}{\end{equation}}
\DeclareMathOperator{\trace}{Tr}      % Trace of matrix

\begin{document}

\title{Emergent percolation length and localization in random elastic networks} %Localization-delocalization/delocalization transition of phonons in disordered solids}

\author{Ariel Amir$^*$}
\affiliation { Department of Physics, Harvard University, Cambridge, MA 02138, USA}
\author{Jacob J. Krich$^*$}
\affiliation {Department of Physics, University of Ottawa, Ottawa, ON, Canada}
\affiliation {Department of Chemistry and Chemical Biology, Harvard University, Cambridge, Massachusetts 02138, USA}
\author{Vincenzo Vitelli}
\affiliation {Instituut-Lorentz for Theoretical Physics, Leiden University, Leiden NL 2333 CA, The Netherlands}
\author{Yuval Oreg}
\author{Yoseph Imry}
\affiliation { Department of Condensed Matter Physics, Weizmann Institute of Science, Rehovot, 76100, Israel}

\begin {abstract}
We study, theoretically and numerically, a minimal model for phonons in a disordered system. For sufficient disorder, the vibrational modes
of this classical system can become Anderson localized, yet this problem has received significantly less attention than its electronic counterpart. We find rich behavior in the localization properties of
the phonons as a function of the density, frequency and the spatial
dimension. We use a percolation analysis to argue for a Debye spectrum
at low frequencies for dimensions higher than one, and for a
localization/delocalization transition (at a critical frequency) above two dimensions. We show
that in contrast to the behavior in electronic systems, the transition
exists for arbitrarily large disorder, albeit with an exponentially
small critical frequency. The structure of the modes reflects a
divergent percolation length that arises from the disorder in the
springs without being explicitly present in the definition of our
model. Within the percolation approach we calculate the speed-of-sound
of the delocalized modes (phonons), which we corroborate with
numerics. We find the critical frequency of the localization
transition at a given density, and find good agreement of these
predictions with numerical results using a recursive Green function
method adapted for this problem. The connection of our results to recent experiments on amorphous solids are discussed.
\end {abstract}
%63.50.+x Vibrational states in disordered systems
%63.20.Pw Localized modes
%72.20.Ee Anderson localization, hopping transport,
%62.30.+d Mechanical and elastic waves; vibrations

 \maketitle

\section{I. Introduction}
Despite decades of experimental and theoretical efforts, the mechanical properties of disordered solids remain poorly understood.  Long-standing challenges such as explaining the low-temperature properties of glasses and ultrasound propagation in granular media require a robust understanding of the effect of disorder on vibrational modes and their localization properties. Interest in phonon localization in disordered solids has been rekindled by recent experiments on colloidal glasses that mimic several vibrational properties of molecular glasses on larger length and time scales \cite{ghosh,chen,kaya}. A significant advantage of colloidal glasses is that their vibrational modes can be experimentally reconstructed, and they are found to exhibit fascinating localization properties. Furthermore, the vibrational modes have been shown to be relevant to understanding the irreversible rearrangements of the system \cite{harrowell,brito,xu,manning1, manning2}.

\begin{figure}
\includegraphics[width=3.375in]{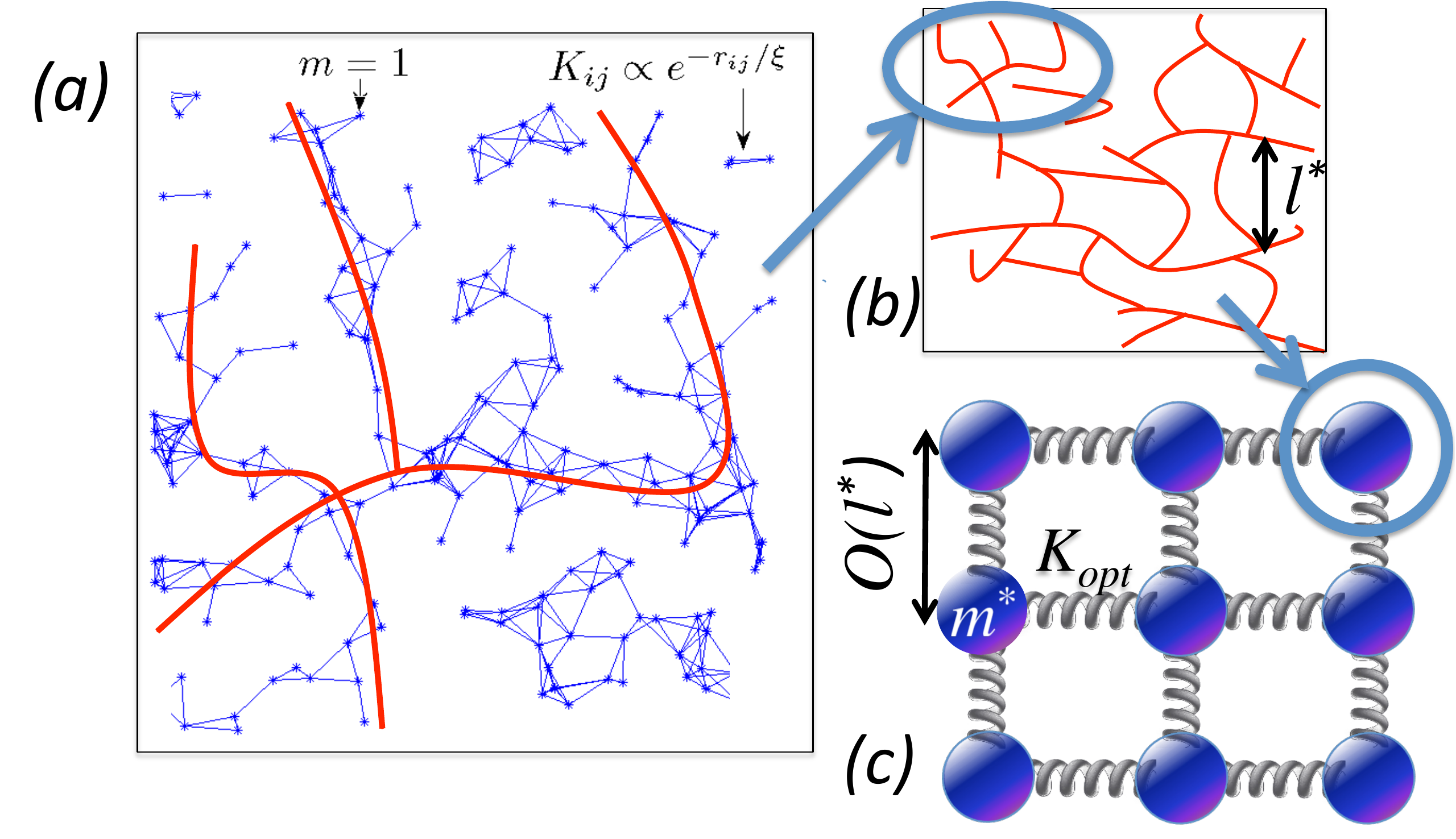}
%\vspace{-0.4cm}
\caption{Illustration of our calculational scheme. (a) Start with a network of identical masses placed randomly, connected by springs with broadly distributed spring constants -- their value exponential in the Euclidean distance between each pair of masses. Only springs with a strength exceeding a threshold $K_{opt}$, defined in the main text, are drawn as thin lines. (b) A  percolating network of connected strong springs is obtained, yielding a meshwork of pathways, with characteristic length scale $l^*$ \cite{efros}. (c) The disordered network is coarse-grained to an effective \emph{ordered} elastic network, with masses $m^*$, spring constants $K_{opt}$ and length scale of order $l^*$. For system size $L \gg l^*$ and dimensions $d >1 $, the low-lying modes follow continuum elasticity, leading to a Debye spectrum. }%The behavior of one and two dimensional systems is different, and the low-lying modes can be localized.}
\label{fig1}
\end{figure}

Since Anderson's seminal work \cite{Anderson}, the localization of electronic wave-functions in disordered metals has been the subject of several successful theoretical investigations \cite{abrahams}. The study of the vibrational modes of a disordered network of masses and springs (see Fig.\ \ref{fig1} ), which is analogous to the electronic problem, received significantly less attention. Formally, the spring problem is different from the electrons problem due to mass conservation, leading to the presence of a zero energy mode -- see Eq. (\ref{model}). Previous approaches to study this problem include perturbative analysis \cite {grigera3} for weakly disordered solids, transfer matrix techniques restricted to effectively one dimension \cite{dyson, alexander, ziman}, numerical studies \cite{simdyankin, roemer, levelstat3d, kottos, vitelli,Ciliberti} and renormalization group approaches \cite{sompolinsky, hastings, Exp_Mat, garel}. One of the successful approaches to modeling vibrations in disordered systems has been using the framework of Euclidean random matrices \cite{distance_matrices, simons, bogomolny, parisi_euclidean, grigera1,grigera2,grigera3,parisi3,skipetrov,rayleigh,euclidean_review}. This problem is mathematically related to the localization problem on sparse graphs \cite{bolle, zippelius, dean,luck}, to the problem of a random walker in a disordered environment \cite{scher,schirmacher,Exp_Mat}, and was also used to explain aging and slow relaxations in electron glasses \cite{amir_glass,amir_aging, amir_review}.

Many important questions remain open regarding the nature of the localization transition of the vibrational modes in disordered systems, its dependence on the dimensionality and the type of the disorder (non-uniform masses versus non-uniform springs, for example).
Most studies rely on numerical solutions or study a particular asymptotic limit, and analytic solutions for tunable disorder and system dimensionality are scarce. In this work, we present a model that is analytically tractable, leading to insights into the way the localization transition depends on the disorder magnitude and dimensionality, and elucidating the emergence of a disorder-dependent length scale which determines various elastic properties of the system.

Continuum elastic theories of disordered solids typically break down below a mesoscopic length scale $\ell^{*}$ (controlled by the amount of disorder) intermediate between particle diameter and system size \cite{wyart,liu,hecke}. In jammed packings of grains and random networks, $\ell^*$ exhibits a power law divergence whenever the network connectivity or packing density are lowered towards the critical values necessary to hold these systems rigid. At low densities (weak connectivity) the effect of disorder is stronger. Plane-wave vibrational modes, which satisfy the Debye density of states, exist only below a critical frequency $\omega^* \sim 1/{\ell^*}$  that goes to zero at the critical point.

We present a minimal model for a random elastic network that is able to capture these properties. The model is presented in detail in section II.
In section III, we argue for Debye spectrum at low frequencies, for dimensions higher than one. Sections IV and V are the core of this work, where we use percolation theory, as shown schematically in Fig.\ \ref{fig1}, to analytically study the nature of the eigenmodes in the case of strong disorder. We find that above two dimensions there is always a localization transition at finite frequency, for arbitrarily large disorder. We find an approximate analytical expression for the critical frequency, which we verify numerically using a recursive Green function and finite-size scaling technique, presented in section VI. For the delocalized modes, we calculate, analytically and numerically, the speed-of-sound, which is directly related to the low-frequency spectrum. The mesoscopic length scale $\ell^{*}$ emerges naturally from the percolation picture, and plays a significant role in the calculation of the speed-of-sound and the phase diagram. In section VII, we discuss the implications of these findings for experiments on colloidal glasses and other amorphous solids.

\section{II. The model}
\label{model_sect}
Our model of the vibrational properties of disordered solids is schematically illustrated in Fig.\ \ref{fig1}a, which shows point particles (with equal masses, $m=1$) randomly and uniformly distributed in a $d$-dimensional space, with average nearest-neighbor distance $r_{nn}$. Between every pair of points $\{i,j\}$ there is a spring with spring constant $K_{ij}$. We choose:
\be
	K_{ij}=\frac{U}{\xi^{2}}e^{-r_{ij}/\xi}, \hspace{0.1 cm}
	K_{ii}=-m\sum_{j\neq i}K_{ij}, \label{model}
\ee
where $r_{ij}$ is the distance between particles $i$ and $j$, $U$ is a characteristic energy scale and $\xi$ is the range of the exponential interaction. The vibrational frequencies $\omega_i$ are related to the eigenvalues $\lambda_i$ of the $N \times N$ matrix $K$, as $\omega_i^2=-\lambda_i$. This model describes scalar elasticity, \emph{i.e.}, it neglects the vectorial nature of the forces \cite{distance_matrices, Exp_Mat}.  The randomly chosen points represent the equilibrium locations for the masses. The dimensionless ratio $\varepsilon \equiv \xi/r_{nn}$ controls the strength of the disorder --  the low density limit ($\varepsilon \ll 1$) corresponds to strong disorder.

The reduction to scalar elasticity is not a controlled approximation, but rather a commonly used simplification \cite{Monthus10a,Chaudhuri10}. Nonetheless, previous works suggest that the essential properties of the system are preserved; for example the boson peak and its scaling properties can be studied within scalar elasticity \cite{scalar1}, while the flat diffusivity leading to a plateau in the thermal conductivity was first predicted within a scalar elastic model \cite{scalar2}.

The sum of every row and column of $K$ vanishes, due to the negative diagonal elements. This `sum-rule' is a result of momentum conservation, which has important consequences for the localization properties and the resulting vibrational spectrum \cite{parshin,pinning}. For example, this constraint makes the matrix negative-semi-definite, which guarantees the system vibrates around a stable equilibrium and leads to a purely delocalized mode with $\omega=0$, corresponding to a uniform translation of all masses.
Without the sum rule, this model is the Lifshitz model for energies in disordered semiconductors \cite{lifshitz}. It is precisely the difference between the electronic and phonon problem that enables us to use the percolation formalism here. As we will discuss, the sum-rule makes the problem mathematically analogous to that of electrons incoherently hopping between sites, for which the percolation approach has been successfully applied \cite{ambegaokar}.

Anderson localization is a wave phenomenon, and as such is expected to occur for the classical phonons studied here, for sufficiently strong disorder. Yet it is not obvious a priori whether the low-frequency vibrational modes of this system should be localized or delocalized; It was previously shown that the states with $\omega>\omega^*$ are localized, with $\omega^*\rightarrow0$ at low density \cite{Exp_Mat}. However, since there is always a delocalized mode at $\omega=0$, from continuity one might expect that in a large enough system, modes of sufficiently low frequency would also be delocalized. We shall show that this is indeed the case above two dimensions. Another, perhaps simpler, question regards the structure of the density-of-states (DOS) of the eigenvalue distribution at low frequencies, related to the diffusive properties of the associated random-walk problem \cite{doron}. Will it be the same as that of an ordered system, manifesting a Debye spectrum, $P(\omega) \sim \frac{\omega^{d-1}}{c^d}$ (with $c$ the speed-of-sound), or can the disorder change it qualitatively? One may think that a Debye spectrum and delocalized modes should go hand-in-hand, but we shall shortly give a counter-example in a one-dimensional system. In dimensions higher than one we show that one always obtains a Debye spectrum at sufficiently low frequencies.

In one dimension and at low densities, the model we study reduces to a chain of equal masses connected by random springs. For low densities we can neglect matrix elements beyond nearest-neighbors, and obtain a power-law distribution of the remaining ones  \cite{alexander,ziman}. For sufficiently small disorder (large enough $\varepsilon$), one obtains a Van-Hove singularity of $P(\omega) = const$ at small eigenvalues, which is Debye-like, while for densities lower than the critical density $\varepsilon = 1$ a sharper power law $P(\omega) \propto  1/\omega^\alpha$ occurs, with $0<\alpha <1$. In the one-dimensional case, all the eigenmodes are localized for any density, with the localization length diverging as $\omega \rightarrow 0$. The fact that the spectrum can still be Debye-like for $\varepsilon>1$ shows that one can have localized states and Debye simultaneously. Note, however, that at high densities the nearest-neighbor model and the one discussed here do not match, since in our case far neighbors will also be important. We now proceed to analyze the model in higher dimensions, focusing mainly on low densities.

\section{III. Density-of-states at low frequencies \label{alternate_deriv}}
We argue that the density-of-states corresponds to a Debye spectrum at low enough frequencies and for $d>1$, for arbitrarily small  $\varepsilon$. To show this, it will be useful to employ two mappings between the vibrational problem studied here and 1) a classical random walker in a disordered landscape \cite{schirmacher, reuveni} and 2) electrical networks \cite{ambegaokar}.

In the random-walker problem, $K$ describes the Markov process, and the probability of being on site $j$ satisfies $\dot{p}_j=K_{ji}p_i$; the sum-rule now enforces probability conservation. For this diffusion problem, the return probability $\psi_{i}(t)$ is defined as the probability to be at site $i$ at time $t$ given that the walker was at site $i$ at time $t=0$. It can be shown that the return probability scales as \cite{Exp_Mat} $\psi(t)\equiv\overline{\psi_i(t)} \propto \int P(\lambda)e^{-\lambda t}d\lambda \label{laplace}$,
where $\psi(t)$ is the average over all sites $i$, and $P(\lambda)$ is the density of states of the matrix $K$.
This formula implies that a higher DOS at small frequencies corresponds
to a larger return probability at long times, which suggests slower diffusion.
In general, a Debye spectrum (in any dimension) corresponds to the
case of normal diffusion and the return probability scales
as $1/{(Dt)}^{d/2}$.
Thus, to establish the existence of a Debye spectrum at low frequencies,
it suffices to prove that at asymptotically long times a particle
will undergo normal diffusion.

To perform this step, we employ the second mapping, between random walks and electrical networks. Consider a random resistor network whose resistance between two sites is proportional to the matrix elements of $K_{ij}$,
\emph{i.e.}, the hopping rate. Then, the Einstein relation tells us that
the diffusion coefficient of the network (if it is non-zero) will
be proportional to the conductivity. Therefore, the question of establishing
normal diffusion is equivalent to proving that a finite conductivity exists in the network. This result was proved using
percolation methods \cite{ambegaokar}, in an identical system where the matrix elements (resistors) between two sites depend on both their energy difference and distance.

For the case of very high (infinite) temperature, the matrix elements depend only
on the distance, exponentially, as in the case we discuss here. Thus, for dimensions higher than one, the result of Ref.\ [\onlinecite{ambegaokar}] shows that there is conductivity, and therefore diffusion, at asymptotic times. Moreover, this allows us to derive
the diffusion coefficient, since percolation theory gives
the dependence of the conductivity on $\varepsilon$: $\sigma\propto D\propto e^{-P_d/\varepsilon}$ \cite{ambegaokar}.

We now explain how this low-frequency Debye-spectrum relates to the results of Ref.\ [\onlinecite{Exp_Mat}], where a strong (non-Debye) peak was shown to occur in the DOS  at very small frequencies. There, it was shown that for low densities ($\varepsilon\ll1$) the DOS can be approximated
by
\be P(\omega)=\frac{dV_{d} e^{-\frac{V_{d}}{2}[\log(\omega^2/2)]{}^{d}}{[\log(\omega^2/2)]}^{d-1}}{\omega}. \label{low_density_dos} \ee
where $V_d$ is the volume of the $d$-dimensional sphere.  This result was derived by approximately calculating all moments of the eigenvalue distribution to an accuracy of $O(\varepsilon^{m})$ with $m>1$.
%Equation (\ref{low_density_dos}) was derived using the statistics of isolated pairs of states and should fail at frequencies where the eigenstates are delocalized.
If the correct DOS is a Debye spectrum at low frequency and Eq.\ (\ref{low_density_dos}) for $\omega>\omega',$ then for sufficiently small $\omega'$ the corrections to the moments will be sufficiently small to be consistent with the previous result. We expect that $\omega'\approx\omega^*$, where $\omega^*$, the delocalization transition frequency described in section V, is sufficiently small.

Although we have shown that above one dimension we expect Debye behavior to set in at sufficiently low frequencies, the existence of Debye spectrum does not prove that the eigenmodes  are delocalized; indeed, the one-dimensional example in section II, at low disorder, has \emph{all} eigenmodes localized while the low frequency spectrum is of the Debye form. What happens in higher dimensions? In Appendix A, we show that for the case of small disorder $\varepsilon \gg 1$, the eigenmode behavior is reminiscent of ordered systems: plane waves are approximate solutions to the eigenvalue problem \cite{distance_matrices} for $ |\vec{k}| r_{nn} \ll 1$ and $\varepsilon \gg 1$, where $k$ is the wavevector. In Appendix A we also calculate the speed-of-sound of these delocalized modes, a result which agrees well with exact numerical diagonalization, as we show in  Fig.\ \ref{speed}. Yet it is not clear whether the delocalized nature of the eigenmodes at low frequencies should carry over to the highly disordered case of $\varepsilon \ll 1$ - clearly the eigenmodes cannot be approximately plane waves as they are in the low disordered case. Nevertheless we  now use a coarse-graining argument to claim that while locally the modes are very different from plane waves, they resemble plane waves when averaged spatially. We use percolation theory to calculate the length scale over which one has to average to achieve this coarse-grained result.

\section{IV. Structure of the eigenmodes: Coarse graining and percolation approach }
\label{percolation_sect}

To determine the low-energy, large-scale modes of the system, we study the response to a static force applied to the outside of a large but finite system; that is, we consider the compressibility. We follow the method for the equivalent problem of random resistor networks \cite{efros}, in which the conductivity corresponds to our model's inverse compressibility. The compressibility of the full system can be well described by considering only a subnetwork of springs, where we keep only $K_{ij}\ge K'$ and set all the other $K_{ij}$ to zero; see Fig.\ \ref{fig1}.  The full network, described by $K'=0$, includes many weak springs, which do not contribute to the rigidity of the system. If $K'$ is large, the subnetwork will not connect the edges of the system, so the compressibility will be infinite. The subnetwork just connects the edges of the system when $K'=K_c=e^{-r_c}$, where $r_c$ is the critical radius of the $d$-dimensional random-site percolation problem \cite{ambegaokar, efros}, $r_c=P_d/\varepsilon$, with $P_d=2(\eta_c/V_d)^{1/d}$, $V_d$ is the volume of the $d$-dimensional unit sphere, and $\eta_c$ is the standard continuum percolation threshold \cite{Lorenz01, Quintanilla07}, giving $P_2=1.20$ and $P_3=0.87$, and we have taken $\xi=U=1$. This subnetwork is still not a good estimate for the full network's compressibility, because the external pressure is concentrated into a large force at a small number of critical links. Reducing $K'$ to a value smaller than $K_c$ gives a large number of parallel rigidity paths, each of which has a critical spring with spring constant approximately equal to $K_c$. This results in a mesh qualitatively represented in Fig.\ \ref{fig1}b. The full network with $K'=0$ will have a compressibility close to that of this \emph{critical subnetwork} with $K'=\alpha K_c\equiv K_{opt}$, for $\alpha$ a constant of order unity, independent of $\varepsilon$ \cite{efros,ledoussal}. Reducing $K'$ further does not significantly change the compressibility, because the weak springs added are `shunted' by the critical subnetwork.

The critical subnetwork has a natural length scale giving the size of its ``holes" (see Fig.\ \ref{fig1}b): the percolation length $l^*=A \varepsilon^{-1}|(r_{opt}-r_c)/r_c|^{-\nu}$, for $r_{opt}=-\log K_{opt}$ close to $r_c$ \cite{efros}, where $A$ is a constant of order unity and the critical exponent $\nu$ is $4/3$ in 2D  \cite{Smirnov01} and $0.875$ in 3D (derived from \cite{Lorenz98}). Since $\alpha$ is $O(1)$, $|r_{opt}-r_c|=|\log \alpha|$ is small, justifying the use of the power-law form: $l^*\sim \varepsilon^{-(\nu+1)}$, which diverges at low density.

All subsystems much larger than $l^*$ have properties (e.g., compressibility) similar to the infinite system, so we can coarse grain the system into boxes with a size of order $l^*$, as illustrated in Fig.\ \ref{fig1}c.
For determining the system compressibility, each box can be approximated as a spring with a spring constant close to $K_{opt}$, corresponding to the critical spring in the network connecting the edges of the box. Due to the exponential spread in $K_{ij}$, all other springs are effectively stiff ($K\gg K_{opt}$) or broken ($K\ll K_{opt}$) \cite{ambegaokar, efros}.  By coarse-graining over regions of order $l^*$, the resulting distribution of critical springs will be narrow, and we have essentially replaced the disordered spring network by an approximately ordered one.

This procedure is only defined in dimensions greater than one, where percolation occurs. For phonons with wavelength large compared to $l^*$, the disorder is averaged away, and the problem maps to that of a \emph{weakly} disordered system; this is the mechanism by which delocalized modes may come about. The lowest non-trivial eigenmodes resemble plane-waves when coarse grained at a scale larger than $l^*$, as shown in Fig. 4 of Appendix B. At smaller lengthscales, however, they are highly affected by the local environment, and appear disordered. At higher frequencies (but still below the localization transition) modes can be hybrids of plane-wave-like mode \cite{parshin}. In the following we use the coarse-graining framework to calculate the speed-of-sound of the plane-wave-like modes and the position of the localization transition.

\section{V. Speed-of-sound and localization transition }
\label{percolation_results_sect}

Using the coarse-graining construction, we find an ordered system on length scales $\gg l^*$, and thus we expect to find large-scale delocalized plane-wave modes. We can readily find the speed-of-sound of these modes at low densities, where percolation applies: $c = l^*\sqrt{K_{opt}/m^*}$, where $m^*$ is the effective mass of each coarse-grained oscillator.  Since the volume of each coarse-grained box is comparable to ${l^*}^d$, we expect the total mass of each block to scale as $m^* \sim \varepsilon^{d}{l^*}^{D_p} \sim \varepsilon^{d-(\nu+1) D_p}$, where the fractal dimension of the percolating cluster $D_p$ is smaller than the system dimensionality $d$, and is approximately $1.9$ in 2D and $2.5$ in 3D \cite{fractal}.. Combining these results for $l^*$, $m^*$ and $K_{opt}$ gives:

\be c \sim e^{-\frac{P_d}{2 \varepsilon}}\varepsilon^{-[d/2+(\nu+1)(1-D_p/2)]}. \label{speed_of_sound} \ee

The framework presented here also also allows us to find where the localization/delocalization transition should occur; The coarse-grained picture cannot describe features smaller than $l^*$, so we expect the delocalized plane-wave modes to cease to exist as their wavelength approaches $l^*$. We can estimate the delocalization transition frequency

\be \omega^*=c k^* \approx \frac{c}{l^*}=e^{-\frac{P_d}{2 \varepsilon}}\varepsilon^{(\nu+1) D_p/2-d/2}.  \label{boundary}\ee

 %We can use the form of $\omega^*$ to calculate the integral of the density-of-states described by Eq.\ (\ref{low_density_dos}), from $\omega^*$ to the upper cutoff $\omega^2=2$, corresponding to a pair of masses \cite{Exp_Mat}. The result it that there is \emph{always} a finite fraction of localized modes.
 Taking the Debye spectrum below $\omega^*$, $P(\omega) \sim \frac{\omega^{d-1}}{c^d}$, and using the value of $c$ from Eq. \ref{speed_of_sound}, we can estimate the number of delocalized modes, and find that it scales as $\varepsilon^{\nu d}$. As one goes to higher disorder (lower $\varepsilon$) the fraction of delocalized modes decreases,  becoming negligible at large disorder.

\section{VI. Numerical analysis of the eigenmodes}
\label{numerical_sect}
\begin{figure}
\includegraphics[width=3.375in]{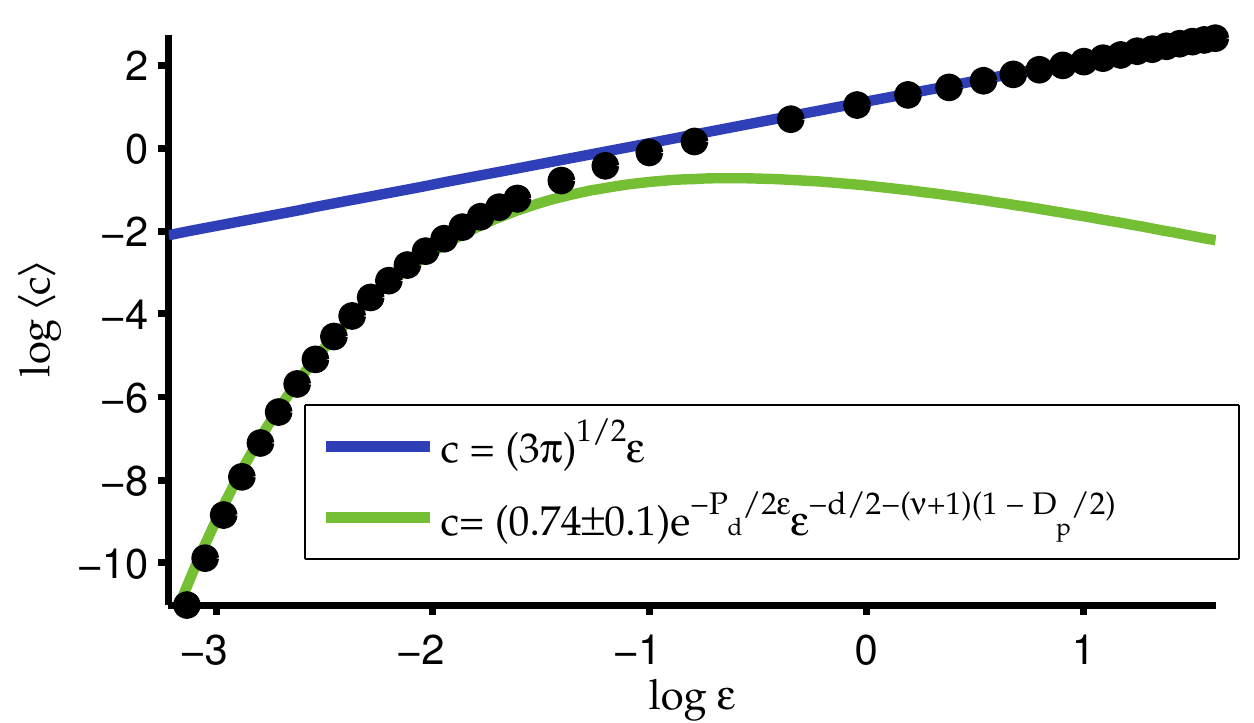}
\caption{The speed of sound of the low frequency delocalized modes as a function of the density parameter $\varepsilon$, for a two-dimensional system. A low value of $\varepsilon$ corresponds to a highly disordered, sparse network, where the distribution of spring constants is very broad. The energy scale $U=1$, see Eq.\ \ref{model}, and $\xi=1$. The straight line shows the exact prediction at high densities, see Appendix A. The other line shows the prediction of Eq.\ \ref{speed_of_sound} of an exponentially suppressed speed at low densities, as predicted from percolation theory, taking the prefactor as a fitting parameter. The points are numerically determined from finite systems, as described in the main text.}
\label{speed}
\end{figure}

We perform numerical studies to test Eqs.\ \ref{speed_of_sound} and \ref{boundary}. The prediction of Eq.\ \ref{speed_of_sound} is corroborated numerically in 2d in Fig.\ \ref{speed}, comparing the exact diagonalization of large matrices. For each $\varepsilon\leq 0.2$, we determined $c$ by averaging results from 2000 realizations with $N=20000$ sites with free boundary conditions. For each $\varepsilon>0.2$ we used a single realization with $N=10000$. For each matrix, the low-lying eigenvectors were determined. We discarded the uniform mode and Lifshitz modes \cite{lifshitz}, where a small number of sites are isolated from their neighbors. The smallest remaining eigenvalue $-\omega_0^2$ gave $c=L \omega_0/\pi$, where the system has linear size $L=N^{1/d}/\varepsilon$, taking $\xi=1$. For $\varepsilon$ from 0.2 to 0.044, the standard deviation in values of $c$ ranged from 1\% to 30\% of the mean, respectively. The points shown in Fig.\ \ref{speed}  are well determined, with statistical error smaller than the symbol size.  Appendix B shows further results that the degeneracies of the low-lying eigenvalues match what is expected from plane waves in an ordered box geometry, see Table S1. Moreover, Fig.\ \ref{planewave} in Appendix B shows the average spatial profile of these low frequency modes, which is plane-wave-like.

Following the same reasoning as the standard one for ordered systems, the DOS which follows from these plane-wave-like modes will be a Debye spectrum, and at low frequencies we find that $P(\omega) \sim \frac{\omega^{d-1}}{c^d}$. This is consistent with the results of the alternative derivation for the form of the DOS at low frequencies presented in section III, %\ref{alternate_deriv},
which relies on mapping this problem to the diffusion problem of a classical random walker in a random landscape.

\begin{figure}
\includegraphics[width=3.375in]{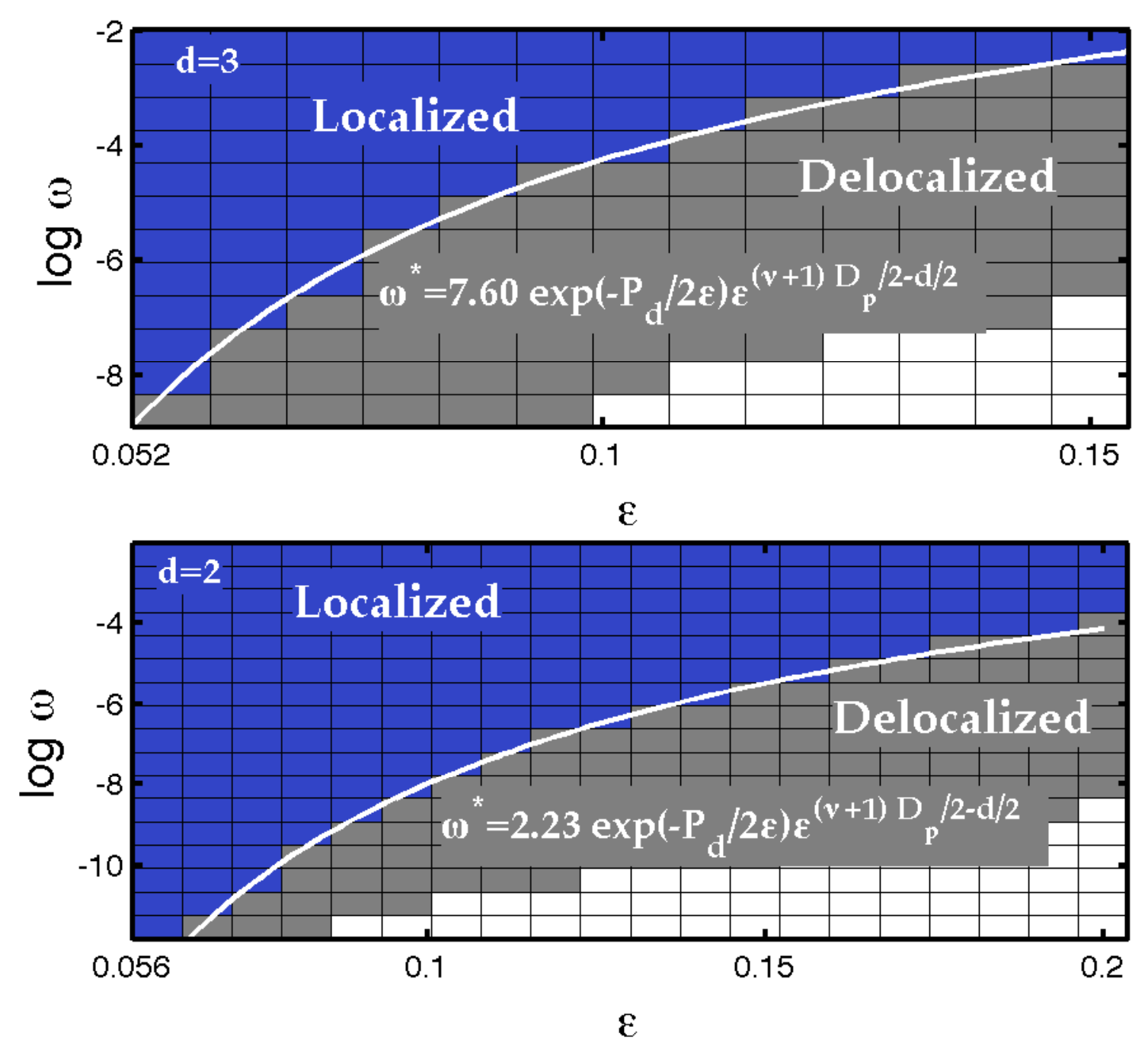} %d2PhaseDiagramJul16A.pdf}%percolation_v4.eps}
\caption{Phase diagram showing the localized and delocalized modes dependence on density ($\varepsilon$) and frequency in two and three dimensions. As predicted in Eq.\ \ref{boundary}, there are both localized and delocalized modes at any density, and at low densities the critical frequency $\omega^*$ below which modes are delocalized depends on $\varepsilon$ according to Eq.\ \ref{boundary}. The only fitting parameter in each line is the prefactor. White squares indicate parameters where states are so strongly delocalized that numerical results did not converge; see Appendix C.}
\label{fig3}
\end{figure}

We have used a recursive Green function and finite-size scaling technique to systematically investigate whether states are localized or delocalized in the infinite-size system \cite{MacKinnon81,Markos06}, with results shown in Fig.~\ref{fig3}. We adapted the technique from the one used in Ref. [~\onlinecite{krich}], which studied the same system without the diagonal terms of $K$; see Appendix C for details. The percolation prediction of Eq.~\ref{boundary} indicates that the delocalized modes should persist to arbitrarily small $\varepsilon$. Due to numerical precision limitations, we cannot study $\varepsilon \leq0.05$ in three dimensions and thus cannot numerically determine if there is a smaller $\varepsilon$ at which all states are localized.
Nonetheless, the prediction of Eq.~\ref{boundary} is in good agreement with the numerically determined boundary, as shown in Fig.\ \ref {fig3}, for both two and three dimensions. The numerical results for 2d also suggest a phase transition, in contrast to the behavior of electronic systems, where it is known that arbitrary small fluctuations will lead to localization in the orthogonal ensemble (which is the appropriate ensemble considering the symmetries of our problem) \cite{Evers08}. Further numerical and analytical study is needed, however, in order to disprove or establish the existence of delocalized eigenmodes in $d=2$.

\section{VII. Connection to amorphous solids and colloidal glasses}
\label{experiment_sect}

Despite its simplicity, our model captures well various features of realistic amorphous solids, such as (a) the transition between Debye and non-Debye behavior in the spectrum (b) the corresponding change in the nature of the eigenmodes (delocalized versus localized) and (c) the dependence of the speed-of-sound of the delocalized modes on the connectivity of the system. For example, the low frequencies  regime of Debye spectrum has proved very difficult to observe experimentally in colloidal glasses \cite{ghosh2}. Our work shows analytically that the extent of this regime can be extremely small for strongly disordered systems but it exists nonetheless. Our parameter $\varepsilon$ describes the degree of connectivity of the system, and our predictions for the dependence of $\omega^*$, the crossover frequency to a Debye spectrum, qualitatively explains the dependence observed in amorphous solids, where $(Z-Z_c)$ plays the role of $\varepsilon$ (with $Z$ the coordination number).

The emergence of $l^*$ from our theory is another important feature which has been found in real amorphous solids, but never connected to percolation theory. Both in experiments and within our model, there is a power-law divergence of $l^*$ as a function of the `sparsity' of the system (described by $(Z-Z_c)$ or by $\varepsilon$). In both cases $\omega^* \sim 1/ l^*$. We show that within our model the speed-of-sound vanishes as $\varepsilon \rightarrow 0$, with direct parallel to the ``softening" experimentally and numerically observed in jammed solids close to $Z_c$ \cite{majmudar, nagel2, durian}. Therefore, our model provides a natural framework to predict the mechanical properties of these fragile solids close to the threshold of vanishing rigidity.

It should be emphasized, though, that this is a simplified model, amenable to analytic treatment -- it should not be expected to predict quantitatively the transition frequencies in real amorphous solids. Rather, it illustrates that there should be such a transition in the character of the vibrational modes and how it depends on the dimensionality and connectivity of the system. For instance, our analytic framework predicts the dependence of the localization properties on dimensionality, showing that $d=2$ is a critical dimension, and that the behavior in 1d is very different from higher dimensions, because there is no percolation in one dimensional systems.

We believe that further insights in the physics of amorphous solids can be gained from studying further this minimal model and compare its predictions for the thermal conductivity and phonon scattering with relevant experimental studies of glasses at low temperatures.

\section{VIII. Conclusions}
We have studied the vibrational spectrum of a disordered system, using analytical arguments relying on percolation theory and a coarse-graining analysis, numerically exact diagonalization, and finite-size scaling.
We find that for $d>1$, there is a critical frequency $\omega^{*}$ vanishing with density, below which a Deybe DOS is observed. This frequency marks a localization-delocalization transition of the vibrational modes. In two dimensions further studies are needed in order to ascertain whether this is a true phase transition.
We analytically account for the speed-of-sound of the delocalized modes, both for the low and high density regimes, using a percolation approach. A lengthscale $l^*$ emerges, which affects the speed-of-sound and the phase transition. This lengthscale appears in various experiments on amorphous solids, and we discuss the applicability of our work to such systems.
%In the future, it would be useful to study the heat transport within this model, and check whether the existence of these delocalized modes leads %to a diverging heat conductivity, as is the case in ordered lattices \cite{vitelli}, which is intimately related to the question of Rayleigh scattering in %these systems \cite{rayleigh}.
In the future, it would be interesting to study heat transport in these systems and to compare results of the minimal model presented here with those of amorphous solids.

\section{Acknowledgments}
 We thank B. I. Halperin, M. Ortu\~no and P. Le Doussal for useful discussions. AA was supported by a Junior Fellowship of the Harvard Society of Fellows. VV  thanks the Feinberg foundation and the Harvard Society of Fellows for their support.

$^*$ These authors contributed equally to this work.
%In the SM we show that for 2d the resulting speed-of-sound is: $c=\frac{d\omega}{dk}=\sqrt{3\pi \varepsilon^2 U}$.

\section{Appendix A: Speed-of-sound at high densities}

At high densities, Ref.\ [\onlinecite{distance_matrices}] gives a simple argument showing that plane waves solutions can be found at high densities. Indeed, the eigenvalue equation is $\sum_j K_{ij} u_j =  \lambda u_i$, which can be written as:
 \begin{align}
\sum_{j \neq i} f(r_{ij}) u_j +K_{ii} u_i =  u_i,
\end{align}
where $f(r)=\exp(-r/\xi)$. Using the fact that $K_{ii}=-\sum_{j\neq i}f(r_{ij})$, and guessing an eigenmode $u_{j}=e^{i\vec{k}\cdot\vec{r_{j}}}$, the eigenvalue equation becomes
\begin{align}
\sum_{j\neq i}f(r_{ij})[e^{i\vec{k}\cdot(\vec{r_{j}}-\vec{r_{i}})}-1]=\lambda.\label{eigenval}
\end{align}
Provided that the sum can be well approximated by an integral, the LHS is independent of $i$, showing that it is indeed an approximate eigenmode. There are two conditions for this to be valid:
1) The points should be dense enough such that the function $f(r)$ is well sampled (\emph{i.e.}, that we can view the sum as a discretized version of the integral). This is equivalent to the condition $\xi\gg r_{nn}$, \emph{i.e.}, $\varepsilon\gg1$.
2) In a similar fashion, the points should be dense enough that the function $e^{i\vec{k}\cdot\vec{r}}$ is well sampled. This leads to the condition $|\vec{k}| \ll 1/r_{nn}$. If conditions (1) and (2) are met, we find that:
\be
\lambda \approx \varepsilon^d \int d\vec{r} f(\vec{r}) [e^{i \vec{k} \cdot \vec{r}} -1]. \label{lambda_eq}
\ee
If we denote the $d$-dimensional Fourier transform of $f(\vec{r})$ by
$F(\vec{k})$, this leads to \cite{distance_matrices}:  $ \lambda=-\omega^2  = \varepsilon^d [F(k)-F(0)]. $  For our exponentially decaying matrix elements, in 2D we find that for long wavelength phonons with $1/k\gg\xi$,
$\omega=ck$, with:
\be c^2 = 3\pi \varepsilon^2 \label{velocity}, \ee
where $U=\xi=1$.
Thus, these long wavelength phonons are indeed acoustic ones, with a well-defined speed of sound $c$. This result is verified numerically in Fig.\ 2.

Repeating the same procedure for 3D, we find:
\be c^2=16 \pi \varepsilon^3.\ee

\section{Appendix B: Numerical verification of delocalized modes}

We use two numerical techniques to verify the existence of delocalized modes in this model: diagonalization of finite systems and a recursive Green function method with finite size scaling. In this section, we describe the former calculations, and in the next section we describe the latter. In both cases, the numerical analysis must truncate Eq.~1, so the elements $K_{ij}$ are set to zero for $r_{ij}>r_{cutoff}$.

For sufficiently large cutoff values (generally 3-5 times $r_c$, the critical percolation radius, described in section \ref{percolation_sect}), the results are independent of the value of the cutoff, as they should be.

For the exact diagonalization, we must discard some low-energy modes. When a single point is further than $r_{cutoff}$ from its nearest neighbor, the resulting eigenspectrum of $K$ contains a mode perfectly localized at that point with frequency zero. Without $r_{cutoff}$, such modes would still exist, with frequencies exponentially small in the distance to other sites; such modes are called Lifshitz modes \cite{lifshitz}. When looking for low-lying modes, we remove from consideration all such Lifshitz modes, defined as those with participation ratios of order unity with vanishingly small frequencies; such modes are also overrepresented in calculations with free boundary conditions, which are the most convenient to use in the degeneracy analysis below. The contribution of these modes to the density-of-states is negligible for a large enough system and for small enough frequencies, since it goes to zero faster than any power-law (note that this relies on the exponential decay of the matrix elements, and for other forms decaying faster with the distance this assertion might be false).

Aside from the Lifshitz singularities, the low frequency modes appear to be delocalized in these finite systems in two and three dimensions. In two dimensions, the lowest two non-trivial frequencies are nearly degenerate, and in three dimensions triply degenerate, as expected from the analysis of eigenmodes in an ordered square lattice geometry with free boundaries. Looking at higher non-trivial eigenfrequencies, their values are consistent with the coarse-grained ordered picture, predicting particular ratios of the low-lying frequencies, as shown in Table 1.
%The degeneracies of the low frequency modes exactly match what is expected from ﬁtting plane waves in an ordered box geometry, see Table 1.

\begin{table}
\begin{tabular}{|c|c|c|c|c|}
\hline
%Index & Eigenvalue - 2D ordered case & Eigenvalue - 2D numerical & Eigenvalue - 3d ordered case & Eigenvalue - 3d numerical\tabularnewline
  Index & $\omega$, 2D ord.& $\omega$, 2D num. & $\omega$, 3d ord. & $\omega$, 3d num. \tabularnewline

\hline
\hline
1 & 1 & 1 & 1 & 1\tabularnewline
\hline
2 & 1 & 1.022 & 1 & 1.014\tabularnewline
\hline
3 & $\sqrt{2}$ & 1.415 & 1 & 1.033\tabularnewline
\hline
4 & 2 & 1.994 & $\sqrt{2}$ & 1.390\tabularnewline
\hline
5 & 2 & 2.004& $\sqrt{2}$ & 1.397\tabularnewline
\hline
6 & $\sqrt{5}$ & 2.222 & $\sqrt{2}$ & 1.426\tabularnewline
\hline
7 & $\sqrt{5}$ &  2.244 & $\sqrt{3}$ & 1.678\tabularnewline
\hline
8 & $\sqrt{8}$ & 2.826 & 2 & 1.950\tabularnewline
\hline
9 & 3 & 3.001 & 2 & 1.991\tabularnewline
\hline
10 & 3 &3.042 & 2 & 2.037\tabularnewline
\hline
\end{tabular}

\caption{Comparison of the frequencies of the low lying modes of the perfectly
ordered case (ord.) and the numerical results (num.) on the disordered case in 2d and 3d, for free boundary conditions. The first non-trivial frequency is normalized to 1. For the two-dimensional case N=500,000, and for the three-dimensional
case for N=100,000. For both cases $\varepsilon=0.1$}
\end{table}

%information in A5e5a and A2e005e0\_1d3c20
%generated by analyze\_q\_5e5.m

\begin{figure}
\includegraphics[width=0.4\textwidth]{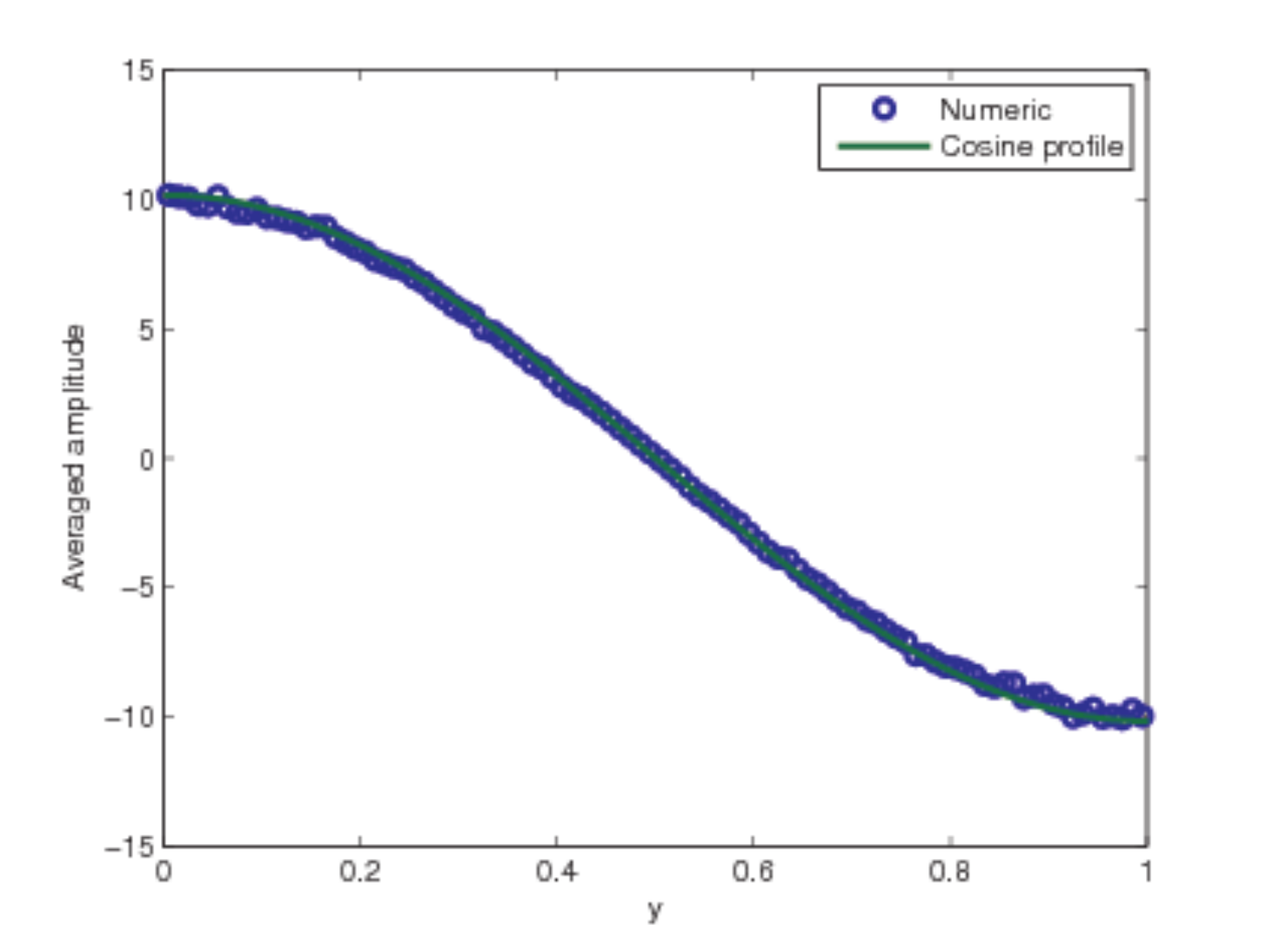}
\caption{The lowest frequency mode is shown for a single matrix with $N=500,000$, $\varepsilon=0.1$, in two dimensions. The amplitudes were binned along one of the sides of the square in which the points are randomly chosen, and the graph shows the sum of amplitudes in each bin. While the eigenmode itself is not a simple plane wave, after `coarse-graining' in this way one obtain a cosine profile, precisely as would be obtained in an ordered system.  \label{planewave}}
\end{figure}

The average spatial profile of these low frequency modes clearly shows plane wave oscillations, as shown in Fig.\ \ref{planewave}.

\section{Appendix C: Recursive Green function calculations}

The recursive Green function technique with finite-size scaling was first used to study Anderson localization by MacKinnon and Kramer \cite{MacKinnon81}. It studies quasi-one-dimensional systems with one very long axis and $d-1$ shorter axes of width $w$.  The system is built recursively by stitching together many $(d-1)$-dimensional strips.  In the original Anderson model, each site is coupled only to its nearest neighbors, so each new slice has direct couplings only to the closest neighbors; this property is essential for the recursion.

In Ref.~ [\onlinecite{Krich11}], we extended this technique to non-lattice problems, as studied in this paper.  Without a lattice, the quasi-one-dimensional system is stitched together from slices of width $w$ and length $l$, which contain a random number of sites, chosen from a Poisson distribution with a mean density $\varepsilon^d$. In order to have particles in each slice interact only with particles in the nearest neighbor slices, we must impose a cutoff on the interaction between sites. That is, we set $\exp(-r)\rightarrow0$ for $r\geq L_c$ for some cutoff $L_c$.
Since we use periodic boundary conditions in the $d-1$ dimensions of the slice, we choose the cutoff to be $L_c=\min[l,w/2]$. Since there is no lattice, $w$ and $l$ must be chosen appropriately large so that the system is not disconnected.

We begin with a single slice of width $w$ and length $l$. We add a second slice and find the portion of the energy-resolved Green function $G(\lambda)$ that connects any site in slice 1 to any site in slice 2. We recursively add more sites, always finding the portion of $G$ connecting sites in slice 1 to sites in slice $N$, $G_{1N}^N(\lambda)=\langle1\vert G^N(\lambda)\vert N\rangle$.  $G_{1N}^N(\lambda)$ is an $M_1 \times M_N$ matrix, where $M_i$ is the number of sites in slice $i$. The probability for a particle injected with energy $\lambda$ into slice 1 to make it to any site in slice N is
\begin{align}
	P_N=\trace[G_{1N}^N(\lambda)]^{1/2}.
\end{align}
Since all states are localized in 1d systems, $P_N$ decays like $\exp(-2 N l/\ell_w)$, for localization length $\ell_w$. We extract the effective localization length at energy $\lambda$ and width $w$ by
\begin{align}
	\frac{2}{\ell_w}=-\lim_{N\rightarrow\infty} \frac{\log P_N}{N l}.
\end{align}
$N$ must be chosen large enough to get good statistical accuracy for $\ell_w$ \cite{Slevin99,Markos06}. Statistically, the system behaves as though it consists of $N l/\ell_w$ independent and identically distributed samples chosen from a normal distribution, so the sampling error decreases as $N^{-1/2}$.

When $\ell_w$ has been acquired for a range of widths $w$, the theory of single-parameter scaling is used to extrapolate to the infinite-size system \cite{MacKinnon81,Markos06,Abrahams79}. According to single-parameter scaling, for sufficiently large $w$ that irrelevant corrections can be neglected, the dimensionless scaling variable $\Lambda_w\equiv \ell_w/w$ obeys a universal scaling law
\begin{align}
	\Lambda_w=F\left[\frac{\phi(\varepsilon,\lambda)}{w}\right]
\end{align}
for some unknown universal function $F$, where $\phi(\varepsilon,\lambda)$ characterizes the infinite-size system: it is the localization length for localized systems and the correlation length for delocalized systems.  In particular, if $\Lambda_w$ increases with $w$, then the state is delocalized and if $\Lambda_w$ decreases with $w$, then the state is localized. This is the approach adopted to determine the phase diagram in Figure 3. Single-parameter scaling has been demonstrated to hold in similar systems without the sum rule \cite{Krich11}, but it has not yet been proved in this system. It is possible that the emergence of the percolation length $l^*$ would cause a breakdown in the single-parameter scaling, but the data so far do not show any such signs.

The details of setting up the recursive Green function calculation have been extensively given elsewhere \cite{Baranger91,MacKinnon83}. The basic principle is that, in each step, the exact Green function $G^N$ of the wire with $N$ slices  and the exact Green function of the new slice $g$ are known. We write $G_0=G^N+g$. The interaction $V$ that connects them can be treated non-perturbatively using a Dyson equation
\begin{align}\label{eq:Dyson}
	G^{N+1}=G_0 + G^{N+1} V G_0 = G_0+G_0 V G^{N+1}.
\end{align}
Matrix elements taken on Eq.~\ref{eq:Dyson} give the recursion relations required.

For our case, the sum rule in the definition of the matrix $K$ means that the interaction V contains not only off-diagonal terms connecting sites in the two slices, but also contains diagonal terms, adjusting the on-site energy of each site.  This does not pose any difficulty for the method, but requires more matrix operations per calculation, increasing the cost of the calculation.

In each calculation, we choose the values of $\lambda$ we wish to study, and for each $\lambda$ store the matrices $G_{1N}^N$, $G_{11}^N$ and $G_{NN}^N$. For the $N+1$st slice being added to the system, it has a Hamiltonian $K$ and the off-diagonal interaction connecting the existing wire to the new slice is $U$. Then in the usual problem, without the sum rule, the result is
\begin{align}
	G_{N+1,N+1}^{N+1}&=(\lambda-K-U^\dag G^N_{N,N} U)^{-1},\\
	G_{1,N+1}^{N+1}&=G_{1,N} U G_{N+1,N+1}^{N+1},\\
	G_{1,1}^{N+1}&=G_{1,1}^N+G^N_{1,N} U (G^{N+1}_{1,N})^\dag,
\end{align}
where we used $g^{-1}=\lambda-K$.
When we include the sum rule, the interaction contains the off-diagonal terms $U$, the diagonal terms $U_N$ acting on the sites in the Nth slice, and the diagonal terms $U_{N+1}$ acting on the sites of the new slice. Then we find
\begin{widetext}
\begin{align}
	G_{N+1,N+1}^{N+1}&=[\lambda-K -U^\dag(\openone_{M_N}-G_{N,N}^N U_N)^{-1} G_{N,N}^N U - U_{N+1}]^{-1},\\
	G_{1,N+1}^{N+1}&=G_{1,N}^N [ \openone_{M_N} + U_N (\openone_{M_N}-G_{N,N}^N U_N)^{-1} G_{N,N}^N] U G_{N+1,N+1}^{N+1},\\
	G_{1,1}^{N+1}&=G_{1,1}^N+\left\{G^N_{1,N} U (G^{N+1}_{1,N})^\dag + G^N_{1,N} U_N (\openone_{M_N}-G_{N,N}^N U_N)^{-1}  [(G^N_{1,N})^\dag+G^N_{N,N} U (G^{N+1}_{1,N+1})^\dag]\right\}.
\end{align}
\end{widetext}

If the calculations are done all at once, there is no need to store $G_{1,1}$. But it is often convenient to break the calculation of a very long wire into many smaller pieces, and then stitch them together. This stitching requires that $G_{1,1}$ be stored.  Consider one wire with $N_1$ slices with recursively calculated Green functions $G^{N_1}_{1,1}$, $G^{N_1}_{1,N_1}$, and $G^{N_1}_{N_1,N_1}$ and a second wire with $N_2$ slices with recursively calculated Green functions $g^{N_2}_{1,1}$, $g^{N_2}_{1,N_2}$, and $g^{N_2}_{N_2,N_2}$. Let $N=N_1+N_2$. Then we can stitch $g$ onto $G$ to find
\begin{widetext}
\begin{align}
	G^{N}_{N,N}&=g^{N_2}_{N_2,N_2}+ G^{N}_{N_1+1,N_2} U g^{N_2}_{1,N_2}+\left\{(g^{N_2}_{1,N_2})^\dag + G^N_{N_1+1,N_2} U g^{N_2}_{1,1}) \left[\openone_{M_{N_1}}-U_{N_1+1} g^{N_2}_{1,1}\right]^{-1} U_{N_1+1} g^{N_2}_{1,N_2}\right\},\\
	G^{N}_{N_1+1,N_2}&=g^{N_2}_{1,N_2}\left[\openone_{N_1+1} - U_{N_1+1} g^{N_2}_{1,1}\right]^{-1} U^\dag G^{N_1}_{N_1,N_1} D,\\
	D&=\left[\openone_{M_{N_1}}-U g^{N_2}_{1,1}(\openone_{M_{N_1+1}}-U_{N_1+1}g^{N_2}_{1,1})^{-1} U^\dag G^{N_1}_{N_1,N_1} - U_{N_1}G^{N_1}_{N_1,N_1}\right]^{-1},\\
	G^{N}_{1,N}&=G^{N_1}_{1,N_1} D U \left\{\openone_{M_{N_1}}+g^{N_2}_{1,1} \left[\openone_{M_{N_1}}-U_{N_1+1} g^{N_2}_{1,1}\right]^{-1} U_{N_1}\right\}g^{N_2}_{1,N_2},\\
	G^{N}_{1,1}&=G^{N_1}_{1,1}+G^{N_1}_{1,N_1} D \left\{U g^{N_2}_{1,1} \left[\openone_{M_{N_1}}-U_{N_1+1} g^{N_2}_{1,1}\right]^{-1} U^\dag + U_{N_1}\right\}G^{N_1}_{1,N_1}.
\end{align}
\end{widetext}

It is generally desirable to let $N$ be large enough that the statistical error in $\ell_w$ is less than $1\%$. For values of $\lambda$, $\varepsilon$ where the states are highly delocalized, $\ell_w$ can be very long, requiring an inordinate length $N$ to get reasonable errors. Conveniently, the values of $\ell_w$ are reasonable for the localized states and the nearby delocalized points, so we can extract the delocalization boundary $\omega^*(\varepsilon)$, while we cannot quite prove that the strongly delocalized states are actually delocalized.
Care must be taken to avoid both overflow and underflow in storing $G_{1,N}$, as detailed in Ref. [~\onlinecite{MacKinnon83}].

\begin{figure}
\includegraphics[width=3.375in]{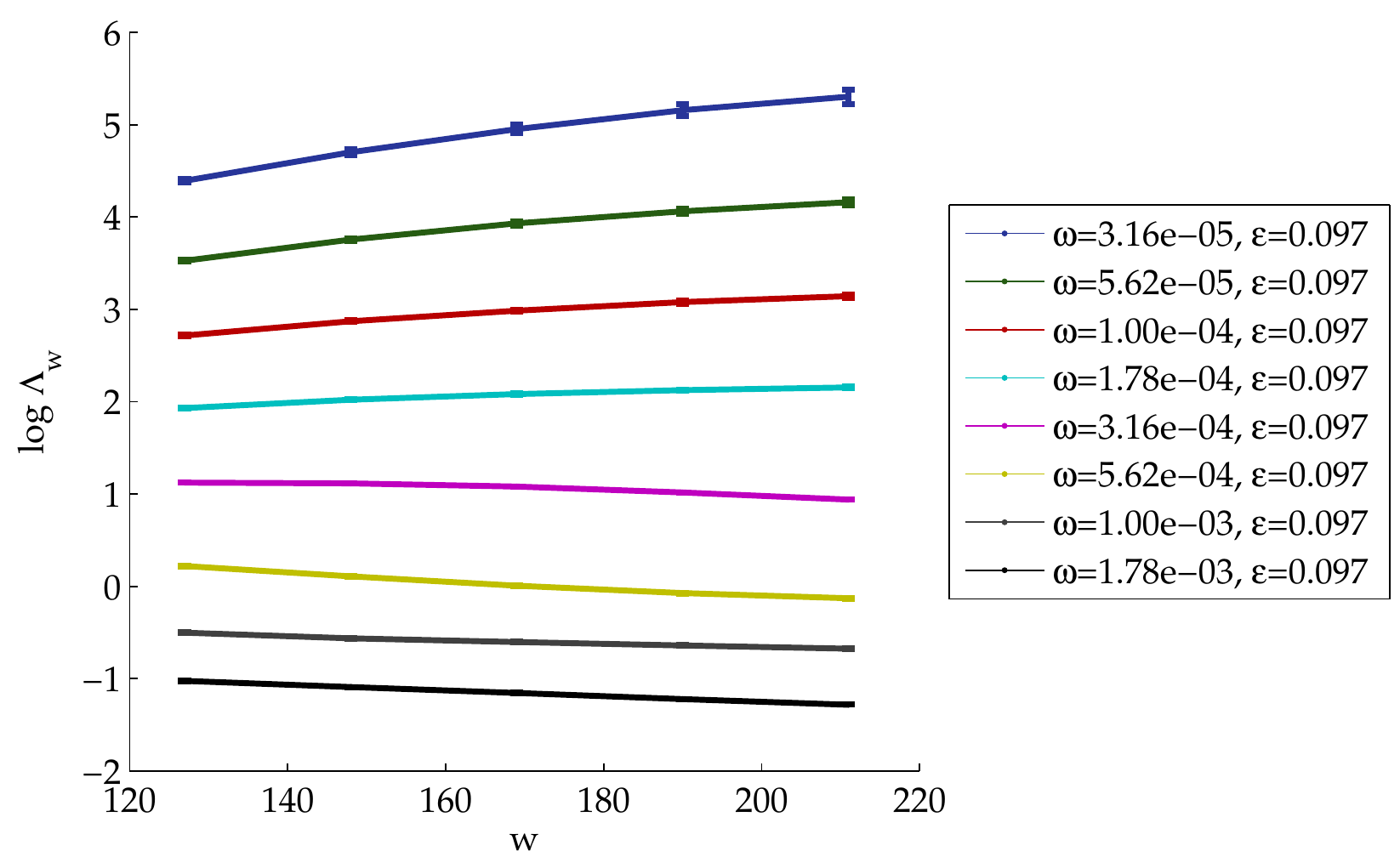}
\caption{\label{fig:rawdata} Part of the data used to produce the phase diagram in Figure 3. For $\varepsilon=0.097$ in $d=2$, we show $\Lambda_w(w)$. The phase boundary between the localized (sloping down) and delocalized (sloping up) energies is clear. }
\end{figure}

Figure \ref{fig:rawdata} shows some of the calculations used to create Figure 3. The statistical error bars are smaller than the data points for all but the lowest energy (most delocalized) state. For these simulations, $l=50$ and $N$ ranged from $2\cdot 10^5$ to $2.6 \cdot 10^7$.
%
%\bibliography{ariel_bib,b-Si}


\end{document}